\documentclass[%
 twocolumn,
superscriptaddress,
preprintnumbers,
 amsmath,amssymb,
 prl,
]{revtex4-1}
\usepackage[dvipsnames]{xcolor}
\usepackage[pagebackref=false]{hyperref}
\hypersetup{
    colorlinks  =   true,
    linkcolor   =   BlueViolet,
    citecolor   =   BlueViolet,
    filecolor   =   BlueViolet,
    urlcolor    =   BlueViolet}
\usepackage{graphics,amssymb,amsmath,epsfig,color}
\usepackage{graphicx}
\usepackage{dcolumn}
\usepackage{bm}
\usepackage{color}
\usepackage{ulem}

\newcommand{\new}[1]{{\color{black}#1}}

\begin{document}

\title{Magnetic field-resilient quantum-limited parametric amplifier} 

\author{Mingrui Xu}
\affiliation{Department of Electrical Engineering, Yale University, New Haven, Connecticut 06520, USA}
\author{Risheng Cheng}
\affiliation{Department of Electrical Engineering, Yale University, New Haven, Connecticut 06520, USA}
\author{Yufeng Wu}
\affiliation{Department of Electrical Engineering, Yale University, New Haven, Connecticut 06520, USA}
\author{Gangqiang Liu}
\affiliation{Department of Applied Physics, Yale University, New Haven, Connecticut 06520, USA}
\author{Hong X. Tang}
\email{hong.tang@yale.edu}
\affiliation{Department of Electrical Engineering, Yale University, New Haven, Connecticut 06520, USA}

\date{\today}

\begin{abstract}
Superconducting parametric amplifiers are crucial components in microwave quantum circuits for enabling quantum-limited signal readout. The best-performing such amplifiers are often based on Josephson junctions, which however are sensitive to magnetic fields. 
Therefore, they require magnetic shields and are not easily integratable with other quantum systems that operates within magnetic fields, such as spin ensemble quantum memories. 
To tackle this challenge, we have developed a  kinetic inductance-based parametric amplifier featuring a NbN nanobridge instead of Josephson Junctions, which provides the desired nonlinearity for a strong parametric gain up to 42\,dB. The added noise of this nanobridge kinetic-inductance parametric amplifier (hereby referred as NKPA) is calibrated and found to be $0.59\pm0.03$ quanta for phase-preserving amplification, approaching the quantum limit of 0.5 quanta. Most importantly, we show that such excellent noise performance is preserved in an in-plane magnetic field up to 427\,mT, the maximum field available in our experiment. This magnetic field-resilient parametric amplifier presents an opportunity towards addressing single electron-spin resonance and more efficient search for Axions as well as Majorana Fermions. 
\end{abstract}

\maketitle
Superconducting parametric amplifiers~\cite{aumentado2020superconducting,Science_JTWPA_2015_Siddiqi,michel_JPA_2010_nature,Lehnert_2008_JPA_nature,vijay_2009_JBA_Review-sci-inst} have proven to be essential in microwave quantum circuits, as they enable detection of extremely weak signals with quantum-limited signal-to-noise ratio. This remarkable technology has been routinely used to read out superconducting qubit states \cite{devoret_2013_SC_outlook,Qubit_JPA_apl_2011_michel} as well as other quantum systems. 
In addition to amplification, parametric amplifiers can be used to generate quantum states, most notably squeezed vacuum state with reduced fluctuation in one quadrature~\cite{Lehnert_2008_JPA_nature}.
The quantum squeezed state is the key to unlocking enhanced detection sensitivity beyond the standard quantum limit~\cite{giovannetti2004_science_standard_quantum_limit} in qubit state readout~\cite{Clerk_2017_enhanced-qubit_squeezing}, nanomechanical resonator motion detection~\cite{Teufel_2016_nature_mechanical_squeezing}, spin-ensemble detection \cite{Squeezing_magnetic_resonance_PRX_2017}, etc.

Despite successful implementations in weak signal detection and noise reduction, the most-popular high-performing superconducting parametric amplifiers are sensitive to magnetic field because of the low critical field of aluminum Josephson junctions~\cite{PRResearch_2019_weides_qubit_magnetic_field}. These amplifiers thus 
have limited compatibility with quantum systems that requires strong magnetic fields.
For example, spin ensembles~\cite{bienfait_2016_ESR_JPA_nat-nano,ESR_JPA_2017_Petta}, quantum dots~\cite{PRApplied_quantum-dot_2015_petta}, dark matter Axions search~\cite{Backes2021-hn} as well as  proposed topological quantum computer~\cite{nayak_2008_RMP_topological-qc} all operate under strong external magnetic fields and could greatly benefit from quantum-limited amplification or squeezed state generation for increased detection sensitivity.
\new{Albeit promising demonstrations, current junction-based technology only allows shielded modular amplifier to be placed outside of the magnetic field \cite{Squeezing_magnetic_resonance_PRX_2017,backes2021NatureAxion}, in which the insertion loss between the source and the amplifier including the connectors, cables and other components becomes a bottleneck of the detection sensitivity, regardless of how close the amplifier's performance is to the quantum limit.
To further improve the detection sensitivity for signals in magnetic field, one pathway is to integrate the amplifier module onto the same chip or system of the source of the signal via a superconducting link, to elimination any excessive attenuation between the signal source and the amplifier.}
Such vision incentivized the development of magnetic field-compatible parametric amplifiers.
Towards this end, various approaches have been pursued to improve the magnetic field resilience of aluminum-based Josephson junctions~\cite{PRApplied_Ando_2022_1T_qubit,PRA_Petersson_2021_magnetic_compatible_qubit}. Moreover, high-kinetic inductance superconducting thin-film such as niobium nitride (NbN) and niobium titanium nitride (NbTiN) became one of the most promising candidates to build magnetic field-compatible parametric amplifiers~\cite{KI_magnetic_field_PRApplied_2016,magnetic_resilient_NbtiN_PRApplied_2019,NbN_magnetic_2021_APL_Romain,GrAl_magnetic_APL_2020} among other material platforms including graphene~\cite{Arxiv_2022_butseraen_graphene_amp,NatComm_2018_magnetic_Graphene,sarkar_2022_graphene_JPA}, granular aluminum~\cite{PRX_2020_Ioan_Granular-aluminum_qubit}, semiconductor nanowires~\cite{Arxiv_Higginbotham_2022_semiconductor-para-amp}, and carbon nanotubes~\cite{PRApplied_2021_Leek_Carbon-nanotube}.  
NbN and NbTiN are known to be resilient to in-plane magnetic fields up to 6~Teslas~\cite{khalifa_2022_magnetic_field_resilient_nanowire,magnetic_resilient_NbtiN_PRApplied_2019,NbN_magnetic_2021_APL_Romain}, and they have the advantages of having a higher superconducting transition temperature (T$_\mathrm{c}$), low fabrication complexity and simple control of nonlinearities.

The past decade has also seen great advances in kinetic inductance-based parametric amplifiers, especially in the form of travelling-wave amplifiers \cite{PRXQuantum_2021_Gao_Travelling-wave-amp_Kinetic-inductance,esposito-2021-perspective-travelling-wave-amp,chaudhuri_2017_kitwpa_2017_apl,ho_Nature-physics_2012_wideband-Kinetic-inductance-amp}, 
many of which demonstrated excellent performance on-par with counterparts built with Josephson junctions. 
The drawback of the travelling-wave amplifier technology is that they typically add noise equal to 3 to 4 times of the quantum limit~\cite{aumentado2020superconducting}, whereas state-of-art cavity-based reflective parametric amplifiers routinely offer noise performances closer to the quantum limit~\cite{michel_JPA_2010_nature,Lehnert_2008_JPA_nature}.
Additional to the noise performance, cavity-based amplifiers also offer advantages including higher gain, suitability for noiseless phase-sensitive amplification and squeezing, as well as less design and fabrication restraints. 
With the goal of pushing the boundary of detection sensitivity for various quantum systems, we focus on developing cavity-based parametric amplifiers using high-kinetic inductance materials that are resilient to magnetic field. 



To make a truly quantum-limited parametric amplifier based on kinetic inductance, a challenge to overcome is to reduce the undesired heating induced by the dissipation of microwave drive power.
Previously reported cavity kinetic-inductance amplifiers made from Nb or NbTiN films \cite{parker_2022_ideal_KIPA,tholen-2007-KI-amplifier-APL} all feature a distributed-element circuit design, leading to relatively weak nonlinearities and large required drive powers on the level of -20~dBm for amplification. 
As a result,  the high drive power dissipation may locally heat up the \new{device and device packaging} and consequently introduce excessive thermal fluctuations to the amplified signal. 
Moreover, given the limited cooling power of a dilution refrigerator, dissipation of the high drive power also prevents scaling up the number of such amplifiers in a single dilution refrigerator unit. 
To overcome these challenges and ensure quantum-limited operation for kinetic-inductance parametric amplifiers, we adopted a weak-link lumped-element circuit structure to drastically increase the nonlinearity.

In this work, we present the design and characterization of a magnetic field-resilient quantum-limited parametric amplifier exploiting the nonlinearity of a NbN nanobridge, which we refer to as nanobridge kinetic-inductance parametric amplifier or NKPA in short. 
Owing to the weak-link design, the NKPA provides a \new{self-Kerr nonlinearity} of $2\pi\times110~\mathrm{kHz}$, 6 orders of magnitudes stronger than previously reported cavity-based kinetic-inductance amplifiers \cite{tholen-2007-KI-amplifier-APL,parker_2022_ideal_KIPA}.
As a result, the drive power requirement for four-wave mixing amplification is reduced to -87\,dBm, which leads to negligible heat dissipation and ultimately ensures quantum-limited operation. 
Phase-preserving amplification over 42\,dB with a gain-bandwidth product of 59~MHz was achieved with the prototype device, and the added noise is found out to be $n_\mathrm{add} = 0.59\pm0.03$ through a noise thermometry calibration, approaching the quantum limit of $0.5$ quanta.
Most importantly, we demonstrate that the excellent NKPA performance is  maintained in in-plane fields up to 427~mT (maximum field provided by the hand-wound coil in this experiment), with the added noise $n_\mathrm{add}$ below $0.68\pm0.07$ quanta for the entire range of the magnetic field. 
Further analysis indicates the potential of magnetic field resilience up to a few Teslas~\cite{magnetic_resilient_NbtiN_PRApplied_2019}.
Finally, we demonstrated phase-sensitive amplification and potential for squeezing with NKPA, leveraging its degenerate-mode configuration. 
Our work represents a significant step towards more robust and more versatile quantum amplifiers. 

\begin{figure}[!h]
\includegraphics{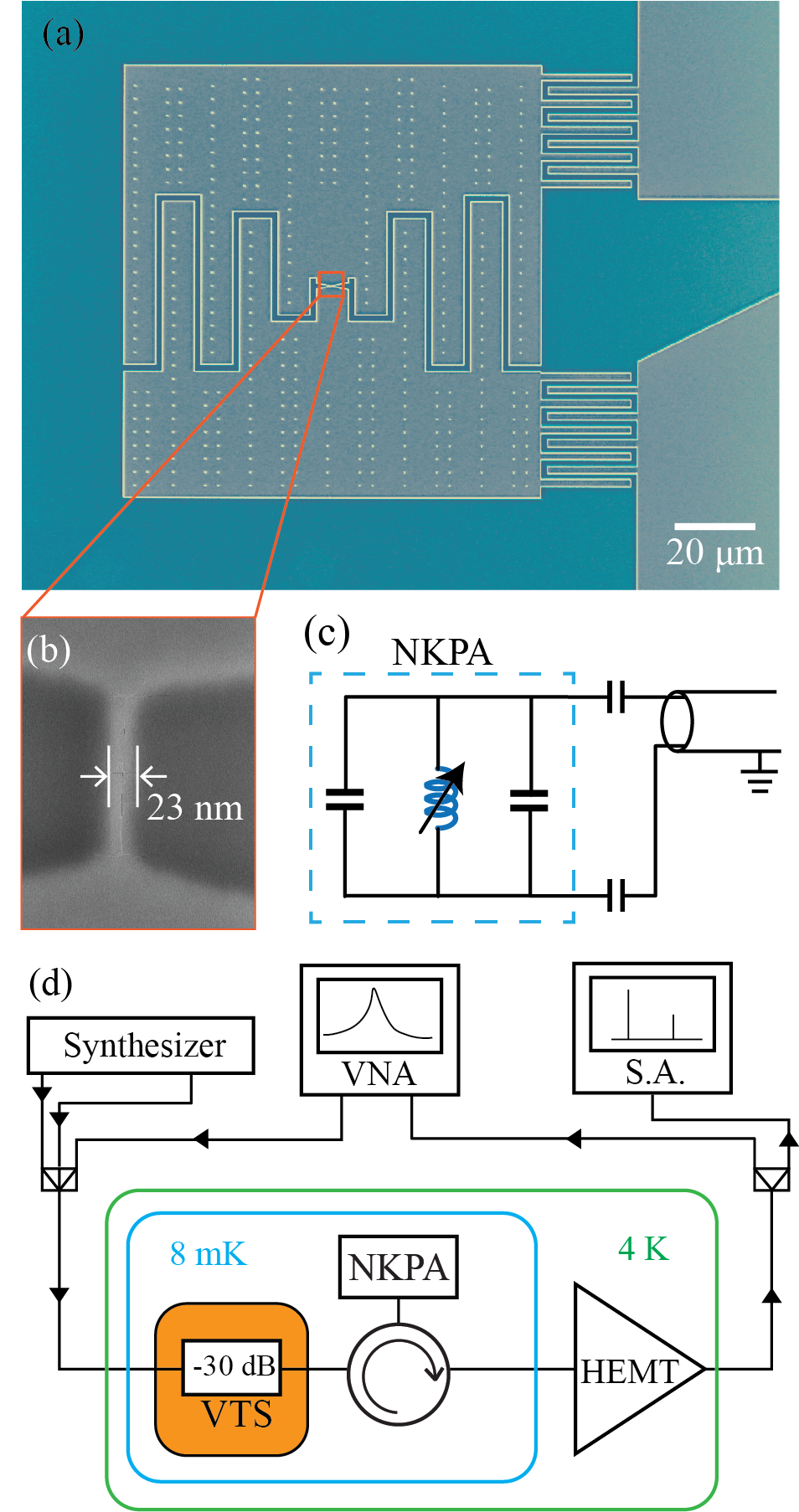}
\caption{\label{fig:diagram} (a) Optical image of the nanobridge kinetic-Inductance parametric amplifier (NKPA). The brighter area is NbN and the darker area is silicon substrate. As illustrated in figure (c), the resonator is constructed as a nonlinear inductor shunted by an interdigitated capacitor. The resonator is then coupled to a microwave transmission line via another set of interdigitated capacitors. (b) is an SEM image of the nanobridge. (d) is the measurement setup diagram, with the NKPA cooled down to 8~mK. The variable temperature stage (VTS) serves as a known thermal source for noise calibration.
VNA and SA stand for vector network analyzer and spectrum analyzer respectively. \new{HEMT stands for High-electron-mobility transistor, which is used as an amplifier.}
}
\end{figure}

The NKPA resonator is constructed as a nanobridge, shunted by an interdigitated capacitor made from NbN thin film, whose optical image and circuit diagram are shown in Fig.~1(a) and 1(c) respectively. On the capacitor is an array of micropatterned holes, otherwise known as ``antidots", which serve a purpose of immobilizing vortices in order to reduce  microwave losses due to vortex motion~\cite{Kleiner_APL_2011_Nb-antidots-magnetic-fields}. On the right side of the NKPA resonator is another set of interdigitated capacitors for coupling the resonator to the transmission line for delivering input signal and pump.
The nonlinear inductance of the NKPA resonator originates in the superconducting nanobridge, see Figure~1(b). In the small current limit, kinetic inductance has a quadratic dependence on the current, typically expressed as $L_\mathrm{k}(I)\approx L_\mathrm{k0}[1+(I/I^*)^2]$, where $I^*$ represents the characteristic current of the nanobridge\textemdash typically on the order of the critical current.  
\new{By discarding all higher order nonlinear terms, for example, $(I/I^*)^4$ , we made an assumption that the current in the nanobridge $I$ is much weaker than $I^*$.}
The quadratic relation results in the third-order (Kerr-type) nonlinearity in the resonator, and is the key to enabling parametric amplifications.

The four-wave mixing Hamiltonian term of a kinetic inductance-dominant LC circuit is $H_\mathrm{int} = K a^\mathrm{\dagger 2}a^\mathrm{2}$, where the vacuum \new{self-Kerr} nonlinearity is expressed as
\begin{equation}
   K= \frac{3}{2}\hbar \omega^3 \frac{\alpha}{Z_\mathrm{0} {I^*}^2}.
\end{equation}
Here $Z_\mathrm{0}$ is the impedance of the resonator, and $\alpha$ is the participation ratio of the nanobridge kinetic inductance in the total inductance of the resonator.
To enhance the nonlinearity, one key advance we made in this work is to drastically shrink the cross-section of the nanobridge \cite{joshi_2022_Nonlinear-nanobridge-TiN,sidiqqi_2009_PRL_Alnanobridge,sidiqqi_2010_APL_Alnanobridge}, and therefore proportionally reduce $I^*$.
Towards this end, we deposit a thin NbN superconducting film on a high-resistivity silicon substrate using atomic layer deposition (ALD) with 90 cycles~\cite{cheng2019superconducting}.
The resulted thickness of the NbN film is approximately 4~nm, covered by a 2-nm thick native niobium oxide layer.
Such a thin NbN film comes with a large sheet kinetic inductance of $179~\mathrm{pH/\square}$. 
Additionally, the film's superconducting critical temperature is around 9 K, which opens opportunity for the device to operate at higher temperatures than the devices based on aluminum Josephson junctions. 
Following the film deposition, we performed a single step of  high-resolution electron-beam lithography with the positive-tone resist CSAR 62 to define the nanobridge with a width of 23~nm and length of 140~nm, as shown in the SEM image in Fig.~1(b). 
Interestingly, comparison between simulations and measurement results suggests the nanobridge inductance is \new{approximately 90\%} larger than the expected value based on the above-mentioned nanobridge size and the superconductor sheet inductance. 
This could be attributed to the approximately 5\,nm ``dead width'' for each side of the superconducting nanobridge, which is caused by the material damage during the etching process as well as the post-etch oxidization of the NbN nanobridge from the edges. These effects help further enhance the nonlinearity, since it reduces the critical current of the nanobridge without significantly compromising the microwave quality factor. 
It is also worth noting that the oxidation process of NbN happens from minutes up to a few days after the film deposition (for surface oxidation of the film) or the film patterning by etching (for side oxidation of the nanobridge). Afterwards, the aging of the device saturates due to the inhibition of further oxidation by the existing niobium oxide layer, protecting the top and side of the NbN nanobridge.  

In addition to the nanobridge cross-section reduction, the resonator geometry is carefully designed to further optimize the nonlinearity by minimizing the impedance $Z_\mathrm{0}$ while maintaining a high kinetic inductance participation $\alpha$. 
Based on Sonnet simulation of the capacitors as well as measured resonant frequencies of NKPA with various nanobridge structures, we extract \new{ $Z_\mathrm{0}=167~\Omega$} and $\alpha = 58.4\%$ for the device with the best nonlinearity. 
The main limit for the participation ratio $\alpha$ is the presence of parasitic inductance of approximately 1.47~nH in total, 
which comes from the interdigitated capacitor (0.73~nH) and the superconducting bank connecting the nanobridge (0.74~nH). Both sources of parasitic inductance could be reduced by adopting thicker capacitors \cite{sidiqqi_2009_PRL_Alnanobridge,sidiqqi_2010_APL_Alnanobridge}.
The geometric inductance is negligible in the nanobridge because of its small size.


\begin{figure}[h!]
\includegraphics{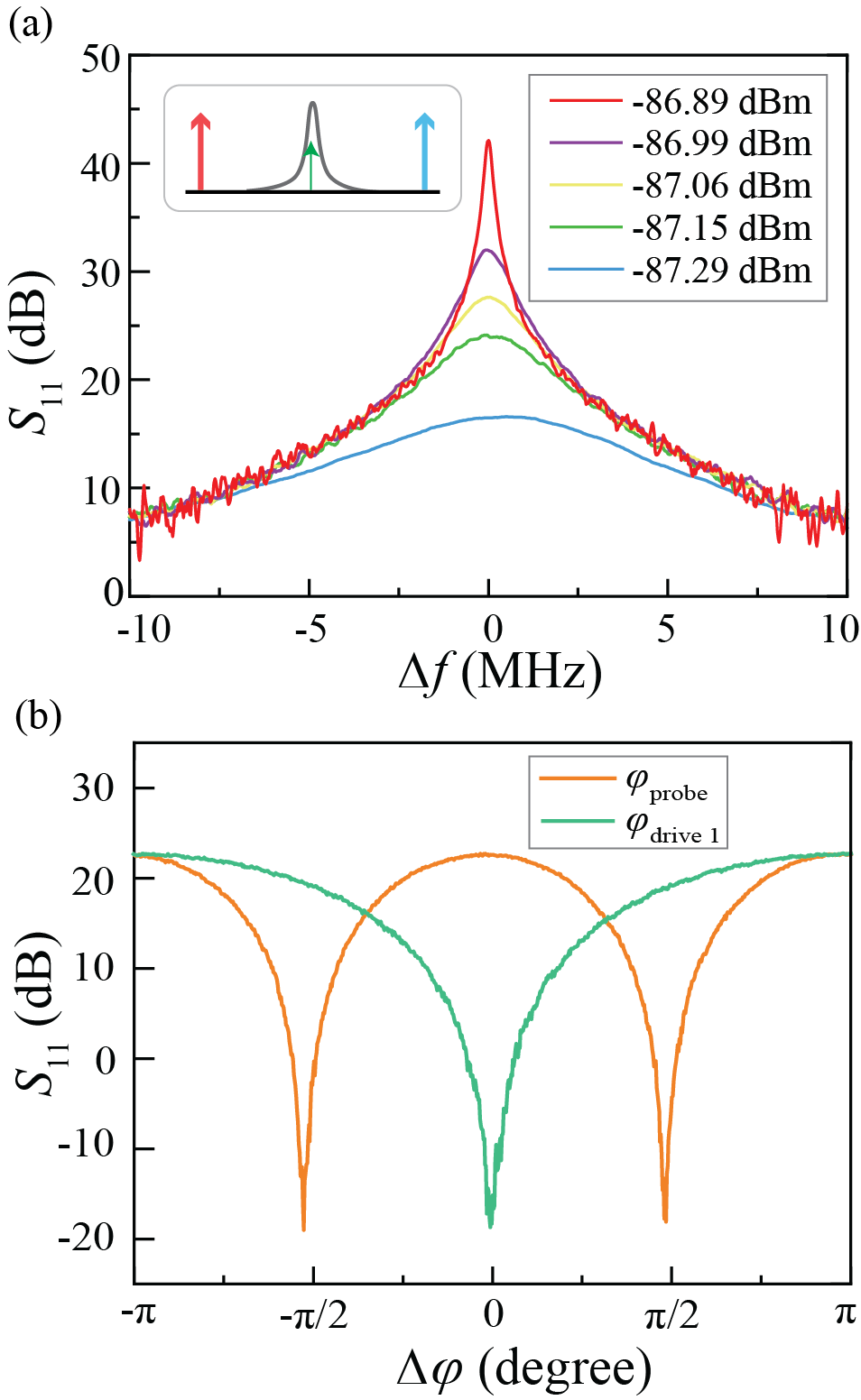}
\caption{\label{fig:diagram} (a) Phase preserving amplification characterized by a reflective vector network analyzer  measurement. To enable gain, we apply a two-tone drive with two frequencies each detuned from the resonant frequency by over 2 linewidths.
(b) Phase-sensitive amplification and deamplification. In this measurement, we send a probe signal at the frequency of $f_\mathrm{s} = (f_\mathrm{p1} + f_\mathrm{p2})/2$. By varying the phase of signal tone and one of the drive tones (data shown for $f_{p1}$), we observe an oscillation between amplification and deamplification with a period of $2\pi$ for $\phi_\mathrm{s}-(\phi_\mathrm{p1}+\phi_\mathrm{p2})/2$.
}
\end{figure}

To characterize the NKPA performance, we package the NKPA chip in an oxygen-free copper housing connectorized with an SMA launcher, and then installed the device in a dilution refrigerator and cooled down to 8\,mK. 
The device is not protected by a magnetic shield as opposed to amplifiers made from Josephson junctions.
The measurement setup is illustrated in Fig.~1(c). 
By sweeping the frequency of a weak probe signal, we obtained the reflection spectrum with a vector network analyzer. 
The low extinction ($< 0.5~\mathrm{dB}$) in amplitude response and $2\pi$ phase shift in phase response indicate
the resonance is in the regime where the intrinsic loss rate is negligible in comparison to the external coupling rate, which is desired for low-noise parametric amplifications.
From the theoretical fitting, we extracted the resonant frequency at 7.45~GHz, the lower bound of intrinsic quality factor $Q_\mathrm{in}$ at 4,000, and a total decay rate of $\kappa_\mathrm{tot}=2\pi \times 58.9~\mathrm{MHz}$, which set the expectation for the gain-bandwidth product for the parametric amplification.


We then drive the NKPA for four-wave mixing parametric amplification. In general, there are two driving schemes. The first is to use a resonant drive tone to enable amplification at frequencies centered at the drive frequency \cite{schuster_2020_PRApplied_FWM_KI}. The second scheme is to use a two-tone drive at frequencies $f_\mathrm{p1}$ and $f_\mathrm{p2}$ each at a side of the resonance, leading to a gain \new{profile} centers at $(f_\mathrm{p1}+f_\mathrm{p2})/2$. 
We adopt the second scheme with each drive detuned by 133.5~MHz \textendash over 2 linewidths, which allows amplified signal analysis without the need to filter out the drive tone.
\new{Importantly, since both drive tones are off-resonant, spurious nonlinear processes which are common in travelling-wave amplifiers \cite{peng_2022_floquet_twpa_spurious_nonlinear_process,TWPA_squeezing_2022_Arxiv_Will-Oliver} are largely suppressed in NKPA. 
For example, numerical simulation confirms that the spurious four-wave mixing process, in which the annihilation of two lower-frequency (or higher-frequency) drive photons generates two photons at the signal and the idler frequencies, can be rightfully neglected.}
By increasing the microwave drive power, we observe rising phase-preserving amplifications as shown in figure~2(a).
Amplification as high as 42\,dB is achieved, demonstrating the advantage of cavity-based amplifier comparing to traveling wave amplifiers, for the absence of spurious nonlinearities. 
At 24~dB gain, the 3-dB bandwidth of the amplification spectrum is 3.1~MHz. 
As a result, the gain-bandwidth product is 74~MHz, close to the expected value equal  cavity linewidth. The 1-dB compression power is found to be -133~dBm, on par with standard Josephson parametric amplifiers consisting of single mixing element \cite{liu_2017_josephson_APL,zhong_2013_squeezing_NJoPhy}. 
In addition, the weak microwave drive power (-87~dBm) for amplification is a testament to the high nonlinearity. Even with a 30\,dB attenuator in the mixing chamber to thermalize the microwave noise for the microwave drive input, the heat dissipation in the dilution fridge mixing chamber is below -56~dBm, which allows scaling the number of NKPA devices without challenging the heat budget. 
With the drive power that generates large gain, we can infer the NKPA \new{self-Kerr nonlinearity $K \approx 2\pi \times 110~\mathrm{kHz}$ and $I^\mathrm{*} \approx 9~\mathrm{\mu A}$}.

As a cavity-based degenerate parametric amplifier, NKPA is also suitable for phase-sensitive amplification and vacuum squeezing. 
To demonstrate the coherent phase-sensitive amplification, we send a probe signal at the frequency of $f_s = (f_{p1} + f_{p2})/2$. 
In this case, the amplified signal and the idler are at the exact same frequency, causing constructive or destructive interference that lead to amplification or deamplification, dictated by the relative phase between the probe and the average phase of the two-tone drive $\mathrm{\Delta} \phi = \phi_\mathrm{probe}-(\phi_\mathrm{p1}+\phi_\mathrm{p1})/2$.
By varying the phase of the probe and the pump, we observe periodic change of the NKPA's gain as shown in Fig.~2(b), with a period of $\pi$ with regard to the probe phase, and a period of $2\pi$ regarding the phase of one of the drives.
This observation matches the expected relations between the gain and the relative phase.
In this measurement, the highest gain is 22\,dB and the deamplification is as deep as 18\,dB, which is limited by the dynamic range of the digitization instrumentation. 
The coherent deamplification demonstration suggests the potential for large vacuum squeezing by NKPA, which could be a unique advantage of NKPA compared to Josephson junction-based amplifiers because of the absence of higher-order nonlinearities~ \cite{Clerk_2017_enhanced-qubit_squeezing,Lehnert_2018_optimal-squeezing_PRApplied,Saddiqi_JPC-higher_order_nonlinearityies_2017_PRApplied,TWPA_squeezing_2022_Arxiv_Will-Oliver}.


\begin{figure}[h!]
\includegraphics{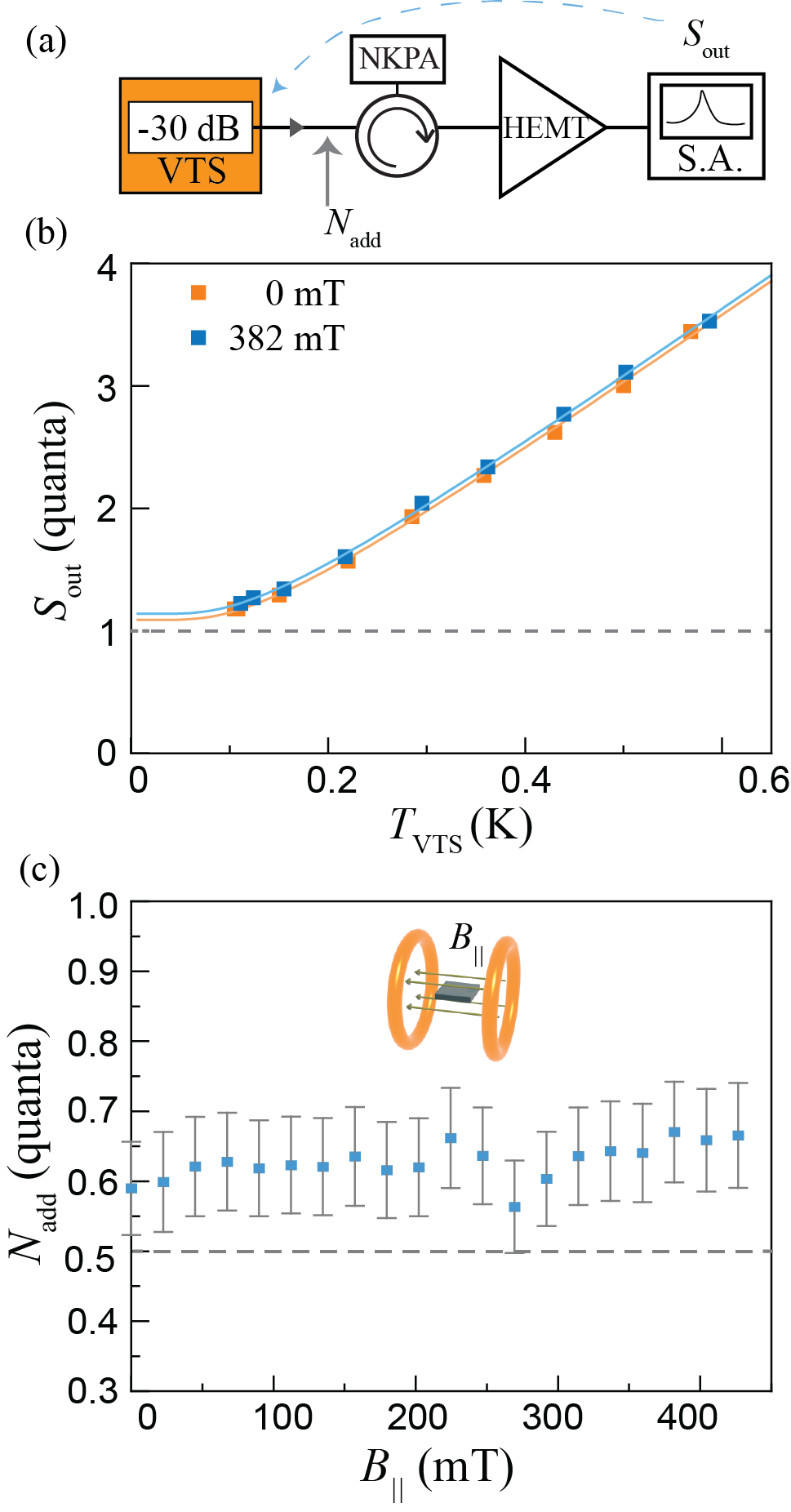}
\caption{\label{fig:diagram} (a) Noise thermometry calibration setup diagram. HEMT stands for high-electron-mobility transistor which is a secondary amplifier. KPA is operated with 26~dB gain during the calibration. The blue dashed arrow indicates that all output noise power spectral density $S_\mathrm{out}$ numbers quoted in this paper are referred back to the output of VTS. (b) Output noise power spectral density $S_\mathrm{out}$ as a function of the thermal source temperature. Dashed horizontal line marks the standard quantum limit of phase-preserving amplification. (c) The added noise $N_\mathrm{add}$ as a function of the in-plane magnetic field $B_\mathrm{||}$.
}
\end{figure}

We then go on to verify the quantum-limited amplification of the NKPA by characterizing its added noise as a phase-preserving amplifier. 
To measure the absolute value of the noise power, we perform noise thermometry calibration with a variable temperature stage (VTS).
By independently varying the physical temperature of a RF load mounted on the VTS, we control the temperature of input noise to NKPA, and then use it as a reference to calibrate the added noise of NKPA.
As shown in the circuit diagram in Fig.~3(a), we use a high-mobility electron transistor (HEMT) amplifier to further boost the signal after NKPA.
During the calibration, we drive the NKPA to provide a 26\,dB gain, to operate in a regime where the added noise of the whole output chain is dominated by the NKPA added noise. 
Then we swept the VTS temperature from 58~mK to 608~mK, and recorded the noise power spectral density at a frequency that is slightly detuned from the center of the gain, which results are delineated in Fig.~3(b) as the orange dots.
Note that in this figure, the y-axis has been normalized to the power spectral density at the output of the VTS.
We then fit the data to a model (solid line) to extract the gain $G$ as well as the added noise $N_\mathrm{add}$ referred back to the output of VTS for the entire output line including NKPA . 
Given the broadband nature of the VTS noise emission, the noise from the VTS at both the signal and the idler frequencies are accounted for (see appendix B).
The data fits the model extremely well, suggesting the validity of the assumptions in the model. 
From the data fitting we extract the system output noise referred back to the output of VTS is $S_\mathrm{out} = 1.09\pm0.03$ quanta, which includes 0.5 quanta vacuum noise and 0.59 quanta added noise, approaching the standard quantum limit of $S_\mathrm{out} = 1$~\cite{Noise_review_RMP_Michel_2010}.

Finally, we study the performance of NKPA in the presence of in-plane magnetic fields. To generate a homogeneous magnetic field, we hand wound a Helmholtz coil using superconducting wires and installed the NKPA device at the center of the coil, as shown in the inset of Fig.~3(c). 
The Helmholtz coil and the device are then mounted on the mixing chamber plate of a dilution refrigerator and cooled to the base temperature of 67~mK.
In this setup, the external magnetic field are perpendicular to the long side of the nanobridge.
\new{Without a second set of balancing coils to adjust the direction of the magnetic field, misalignment between the magnetic field and the device plane is inevitable, which is estimated to be smaller than $1^{\circ}$.}
Upon increasing the bias current of the coil, we can generate a magnetic field up to 427~mT before the superconducting coil quenches. 
The NKPA resonance shifts by 26~MHz under maximum field, while the total linewidth changes by less than 3\% by fitting the reflection spectrum to a input-output theory model with 1\% fitting error~\cite{xu2019frequency}. 
This indicates negligible impact of magnetic field-induced microwave losses.\par

We then characterize the added noise of NKPA in magnetic fields by monitoring the output noise power spectra and subsequently infer added noise using previously calibrated gain of the output line amplification chain.
To maintain the consistency of NKPA gain while sweeping the magnetic field, the two-tone drive power is adjusted by up to 0.3\,dB to compensate for the resonant frequency shift in presence of the external magnetic field.
Results plotted in Fig.~3(c) shows that the added noise remained under $0.68\pm0.07$\,quanta for in-plane magnetic field up to 427\,mT.
As an additional independent verification of the NKPA performance in magnetic field, we performed VTS noise thermometry again with 382~mT in-plane field applied to the NKPA, which results are shown in Fig.~3(b) as the blue dots. 
Theoretical fit (solid blue curve) suggests $N_\mathrm{add} = 0.64\pm0.03$, in excellent agreement with the results in Fig.~3(b). 
Although not surprising, these evidences strongly conclude the robustness of NKPA as a quantum-limited amplifier in external magnetic fields.
The magnetic field resilience could be attributed to the negligible magnetic field-induced loss compared to the external coupling rate. 
Looking forward, NKPA's resilience to in-plane fields up to 6~Tesla remains promising, as intrinsic quality factor of a NbN resonator of $10^4$ has been reported in such a strong magnetic field~\cite{NbN_magnetic_2021_APL_Romain}.

Beyond the prototype device reported above, the unique design of NbN kinetic-inductance nanobridge amplifier offers great engineering flexibility for various applications.
To begin with, the nanobridge characteristic current $I^*$ could be engineered between $2~\mu A$ and 10~mA by simply adjusting the width and thickness of the nanobridge.
As a result, the nonlinearity could vary from 100~kHz to 0.01~Hz.
The high nonlinearity regime where the above-mentioned device resides benefits applications demanding extremely weak signal detection, broad band amplification as well as drive power reduction. 
On the other hand, for experiments with weak signals of high dynamic range, the 1-dB compression power of NKPA could be greatly improved by reducing the nonlinearity with increased width of the nanobridge ~\cite{parker_2022_ideal_KIPA,chaudhuri_2017_kitwpa_2017_apl}.
Moreover, higher gain-bandwidth product could be achieved by designing stronger coupling capacitor to the coplanar waveguide transmission line on-chip.
In addition, as demonstrated in previous works, frequency tunability up to 200~MHz could be achieved by directly feeding a bias current to the nanobridge~\cite{vissers2015_tunable-sc-resonator_apl,2010_nanoinductor_probe}, or inducing a screening current in the nanobridge through a loop structure via an external magnetic flux~\cite{xu2019frequency}. 
Such frequency tunability will largely ease the difficulty to align the frequencies of the signal and the amplifier~\cite{Qubit_JPA_apl_2011_michel}.
Finally, the bias current in the nanobridge could also enable three-wave mixing parametric amplification~\cite{parker_2022_ideal_KIPA,PRXQuantum_2021_Gao_Travelling-wave-amp_Kinetic-inductance}. 


In summary, we introduced a novel nanobridge kinetic-inductance parametric amplifier that is compatible with magnetic fields. The unique weak link design enhanced the nonlinearity by 6 orders of magnitudes comparing to previous reported devices, allowing quantum limited parametric amplification with a weak drive power on par with Josephson parametric amplifiers \cite{michel_JPA_2010_nature,Lehnert_2008_JPA_nature,vijay_2009_JBA_Review-sci-inst}. We also demonstrated that such excellent performance is maintained in the presence of in-plane magnetic fields up to 427\,mT. In the near future, NKPA has the potential to be integrated with spin-ensembles and other quantum experiments requiring magnetic fields to unlock unprecedented detection sensitivities.

\section{Acknowledgments}
\begin{acknowledgments}
The authors would like to thank Professor Michel Devoret, Professor Debdeep Jena, Dr. Alessandro Miano, Wei Dai, and Dr. Xu Han for useful discussions, as well as Dr. Yong Sun, Kelly Woods, Sean Rinehart, and Dr. Michael Rooks for assistance in device fabrication. We acknowledge funding support from Office of Naval Research on the development of nitride-based superconductors (under grant number N00014-20-1-2126) and from Army Research Office on the quantum transducer development (through grant number No. W911NF-18-1-0020). The part of the research that involve cryogenic instrumentation was supported by the US Department of Energy Co-design Center for Quantum Advantage (C2QA) under contract No. DE-SC0012704.  
\end{acknowledgments}

\section{Appendix A, nonlinear resonator model}
The NKPA is constructed as an LC resonator with a highly nonlinear kinetic inductance from the nanobridge. 
Its interaction Hamiltonian can be expressed as:
\new{
\begin{equation}
\label{meta hamiltonion}
\frac{H_\mathrm{int}}{\hbar}=-\frac{1}{4\hbar}\frac{L_{k0}}{I^{*2}}I^4_\mathrm{zpf}(a+a^{\dagger})^{4},
\end{equation}
}
where $a$ is the bosonic operator, $L_\mathrm{K0}$ is the kinetic inductance of the nanobridge, $I_\mathrm{ZPF}$ is the zero-point current of the resonator, and $I^*$ is the characteristic current of the nanobridge. For detailed derivation please see references \cite{yurke-2006-Kinetic-inductane_amplifier,parker_2022_ideal_KIPA}.

To operate the NKPA as an amplifier, we apply two drive tones each at a frequency of $\omega_\mathrm{p1}$ and $\omega_\mathrm{p2}$. We then denote the intracavity amplitude for each drive as $B$ ($C$) and the phase as $\phi_\mathrm{1}$ ($\phi_\mathrm{2}$). For simplicity, we assume these are classical drives. To include these drives, the NKPA intracavity field $a$ expressed in the rotating frame of $(\omega_\mathrm{p1}+\omega_\mathrm{p2})/2$ is
\begin{equation}
\label{a expansion}
a = Be^{i(\Delta t+\phi_\mathrm{1})}+Ce^{i(-\Delta t+\phi_\mathrm{2})}+\delta a,
\end{equation}
where $\Delta = (\omega_\mathrm{p1}-\omega_\mathrm{p2})/2$.


After plugging Eq.~\eqref{a expansion} into Eq.~\eqref{meta hamiltonion} and performing rotating-wave approximation, we arrive at the following Hamiltonian:
\new{
\begin{align}
\begin{split}
\label{kerr hamiltonion}
\frac{H_\mathrm{int}}{\hbar}&=-\frac{K}{4}\delta a^{2}\delta a^{\dagger2}+K(B^2+C^2)\delta a \delta a^{\dagger}\\
&+\frac{K}{2}(BC e^{i(\phi_\mathrm{1}+\phi_\mathrm{2})} \delta a^{\dagger2}+BCe^{-i(\phi_\mathrm{1}+\phi_\mathrm{2})} \delta a^{2}),
\end{split}
\end{align}
}
where the vacuum Kerr nonlinearity is defined as:\new{
\begin{equation}
\label{Kerr original definition}
K=\frac{6}{\hbar}\frac{L_{k0}}{I^{*}{}^{2}}I^4_\mathrm{zpf}.
\end{equation}
}
In Eq.~\eqref{kerr hamiltonion} the first term refers to self-Kerr effect which is negligible when $\delta a$ is a weak signal. The next term refer to cross-Kerr frequency shift, and the last two terms
are responsible to parametric down conversion. Note that the parametric down conversion coefficient $\epsilon = KBC e^{i(\phi_\mathrm{1}+\phi_\mathrm{2})}$ scales with the geometric average of the power of the two drive tones.

Further, to simplify Eq.~\eqref{Kerr original definition}. We can express $I_\mathrm{zpf}$ in terms of lumped element circuit parameters\new{
\begin{equation}
    I_\mathrm{zpf} = \sqrt{\dfrac{\alpha \hbar\omega}{2L_\mathrm{K0}}} = \omega \sqrt{\dfrac{ \hbar }{2Z_\mathrm{0}}}.
\end{equation}
}
Then we can arrive at 
\begin{equation}
   K= \frac{3}{2}\hbar \omega^3 \frac{\alpha}{Z_\mathrm{0} {I^*}^2},
\end{equation}
where the resonator impedance is defined as $Z_\mathrm{0} = \sqrt{\frac{L_\mathrm{K0}}{\alpha C_0}}$.

\section{Appendix B, Noise thermometry calibration}
\new{
To characterize the NKPA added noise, we measure the output noise power $P$ as a function of the variable temperature stage temperature $T_\mathrm{VTS}$, which is expressed as
\begin{equation}
P = BW\cdot G_\mathrm{sys} \cdot S_\mathrm{out}.
\end{equation}
Here, BW is the spectrum analyzer detection bandwidth and $G_\mathrm{sys}$ is the total amplification of the output line including NKPA. Note that $G_\mathrm{sys}$ counts the amplification of the entire output chain from the otput of VTS to spectrum analyzer. As a result, $S_\mathrm{out}$ which denotes the output noise power spectral density is referred to the output of VTS. We could break down $S_\mathrm{out}$ into several components:
\begin{equation}
\begin{aligned}
S_\mathrm{out} & = \hbar \omega_\mathrm{s}(n_\mathrm{VTS,s}  +n_\mathrm{VTS,i} +2n_\mathrm{q} + n_\mathrm{exc}).
\label{sout_expression}
\end{aligned}
\end{equation}

In this equation, $n_\mathrm{VTS,s}$ and $n_\mathrm{VTS,i}$ denotes the thermal occupancy of the VTS at the signal and idler frequencies, which follows the Boltzmann-Einstein distribution $n_\mathrm{VTS,s} = 1/(\mathrm{exp}(\hbar\omega_\mathrm{s}/k_\mathrm{B}T_\mathrm{VTS})-1)$ and  $n_\mathrm{VTS,i} = 1/(\mathrm{exp}(\hbar\omega_\mathrm{i}/k_\mathrm{B}T_\mathrm{VTS})-1)$.
Additionally, we define $n_\mathrm{q} = 0.5$ as the quantum fluctuation, and  $n_\mathrm{exc}$ as the excessive noise of the amplifier. Furthermore, $\omega_\mathrm{s}$ is the signal frequency, and $\omega_\mathrm{i} = \omega_\mathrm{p1}+\omega_\mathrm{p1}-\omega_\mathrm{s}$ is the idler frequency of the parametric amplification.
Note that by writing eq.~\eqref{sout_expression}, we assume large amplification of NKPA, so that the contributions of noise from the signal and idler frequencies are almost identical.\par

In the context of a phase-preserving amplification, the added noise is conventionally defined as $n_\mathrm{add} = n_\mathrm{q}+ n_\mathrm{exc}$, which includes the quantum fluctuation of the idler frequency and any excessive noise. Therefore, the $n_\mathrm{add}$ is extracted by measuring the output noise power $P$ as a function of the $T_\mathrm{VTS}$ and then fit the experimental data to equation
\begin{equation}
P = BW\cdot G_\mathrm{sys}\cdot \hbar \omega_\mathrm{s} (n_\mathrm{VTS,s} +n_\mathrm{VTS,i} +n_\mathrm{q} +n_\mathrm{add}).
\end{equation}
}

It is worth noting that the $n_\mathrm{add}$ is not exactly NKPA's added noise $n_\mathrm{NKPA}$, albeit the proximity. \new{Because the reference plane for $n_\mathrm{add}$ is the output of VTS, while the reference plane for $n_\mathrm{NKPA}$ should be the input of NKPA. There is usually a small attenuation between VTS and NKPA, leading to overestimation of $n_\mathrm{NKPA}$ by approximating it to $n_\mathrm{add}$.} If we follow a simple model where the attenuation between VTS and NKPA is denoted as $\lambda$ and the added noise caused by the rest of the amplification chain including HEMT is $n_\mathrm{sys}$ \cite{parker_2022_ideal_KIPA}. Then the relation between $n_\mathrm{add}$ and $n_\mathrm{NKPA}$ becomes:
\begin{equation}
n_\mathrm{add} = \frac{1}{\lambda}(n_\mathrm{NKPA}+n_\mathrm{sys}/G_\mathrm{NKPA}+n_\mathrm{q})-n_\mathrm{q}.
\end{equation}
\new{To estimate $n_\mathrm{NKPA}$, we plug in measurement results $G_\mathrm{NKPA} = 26~\mathrm{dB}$ and $n_\mathrm{sys} = 23\pm2$, We also estimated that $\lambda = 93\%\pm4\% \: (-0.3\pm0.2~\mathrm{dB})$, citing the insertion loss of circulator QCY-G0401201AU and previous calibration results in reference \cite{PRL_radiative_cooling_me_2020}. Then we arrive at an estimation of the true NKPA added noise on the level of $n_\mathrm{NKPA} \approx 0.50^{+0.03}_{-0.00}$}.

\bibliographystyle{apsrev4-1}
\bibliography{lib.bib}

\end{document}